\documentclass[%
 reprint,
 amsmath,amssymb,
 longbibliography,
 aps,
 pre,
]{revtex4-1}

\usepackage{graphicx}
\usepackage{calc}
\usepackage[normalem]{ulem}
\usepackage{graphicx}
\usepackage{bm}
\usepackage{xcolor}
\usepackage{hyperref}
\usepackage{amsmath}
\usepackage{amssymb}
\usepackage[utf8]{inputenc}
\usepackage{ulem}
\usepackage{xr}
\usepackage{outlines}
\usepackage{bbm}
\usepackage{mathtools}
\usepackage{graphicx}
\usepackage{hhline}
\usepackage{mathrsfs}
\usepackage[normalem]{ulem}
\usepackage{float} 
\usepackage{tikz}
\usetikzlibrary{arrows}
\usepackage{physics}
\usepackage{soul}
\usepackage{bm}
\usepackage{multirow}

\usepackage{subfloat}

\def\Z{\mathbb{Z}}

\def\pMeas{\mathbb{P}^{\bm{\xi}}\left(|\vec{S}(t)|\geq r\right)}

\begin{document}

\title{Universal Fluctuations in the Tail Probability for $d=2$ Random Walks in Space-Time Random Environments}
\author{Franscesca Ark$^*$, Jacob B. Hass$^*$, Eric I. Corwin$^*$}
\affiliation{$^*$Department of Physics and Materials Science Institute, University of Oregon, Eugene, Oregon 97403, USA.}
\date{\today}

\begin{abstract}
    Many diffusive systems involve correlated random walkers due to a shared environment. Such systems can be modeled as random walks in random environments (RWRE). These models differ from classical diffusion in the behavior of the extremes---the walkers that move the fastest or farthest. In spatial dimension $d=1$ RWRE models have been well studied numerically and analytically and exhibit universal behavior in the Kardar-Parisi-Zhang universality class. Here, we study discrete lattice RWRE models in $d=2$. We find that the tail probability exhibits a different universal scaling form, which is nevertheless characterized by the same coefficient, $\lambda_\mathrm{ext}$, as in the $d=1$ case. We observe a critical scaling regime for fluctuations in the tail probability at positions that scale linearly in time.
\end{abstract}

\maketitle

\begin{figure}
    \centering
    \includegraphics[width=\columnwidth]{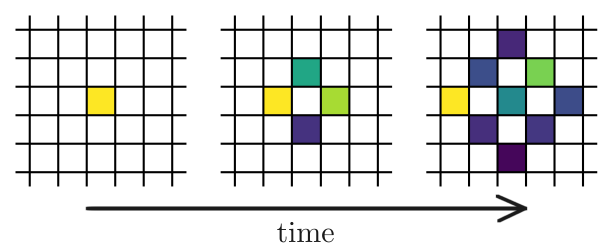}
    \includegraphics[width=\columnwidth]{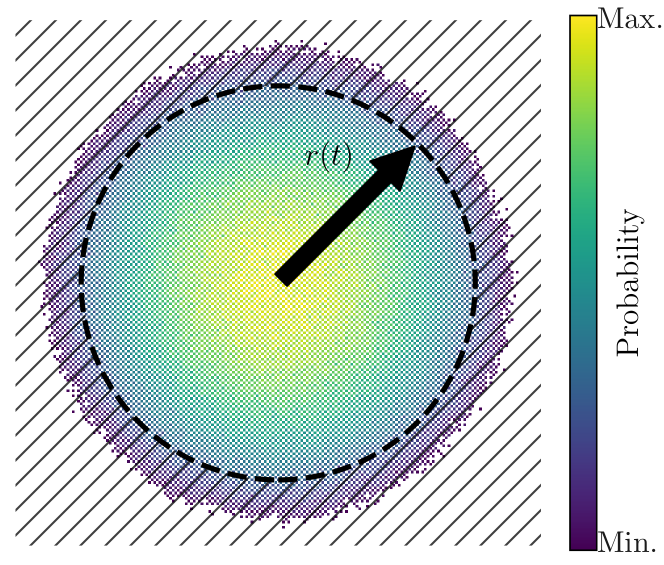}
    \caption{Illustration of the space-time evolution of the probability mass function defined in Eqn.~\eqref{eqn:transitionProb} and the definition of the tail probability $\pMeas$ for an RWRE in $d=2$. The color indicates the relative probability. Top: Probability mass function for $t=0,1,2$. Bottom: Probability mass function at time $t=500$ for a single $\bm{\xi}$. We sum the probability past a circle of radius $r$ (hatched area) to find the tail probability $\pMeas$. }
    \label{fig:model}
\end{figure}

\textit{Introduction---} The random walk in a random environment (RWRE) model takes into account the shared environment in which particles diffuse, unlike classical diffusion which treats particles as independent and identical~\cite{einstein_uber_1905,einstein_zur_1906, einstein_theoretische_1907}. The shared environment is important to the behavior of the extremes: those that travel the fastest or farthest. In many biological and physical systems, such as sperm cells searching for eggs~\cite{reynaud_why_2015, lawley_distribution_2020}, calcium ions traveling across dendritic spinal structures~\cite{basnayake_fast_2019}, and contagions spreading through a population~\cite{keeling_dynamics_2001, hufnagel_forecast_2004,iannelli_effective_2017}, the meaningful action is accomplished by the first, or first handful, of arriving agent(s). Analytical and numerical results show that RWRE models in dimension $d=1$ fall into the Kardar-Parisi-Zhang (KPZ) universality class~\cite{barraquand_random-walk_2017, le_doussal_diffusion_2017, barraquand_moderate_2020, hass_anomalous_2023, hass_first-passage_2024,parekh_hierarchy_2024,das_multiplicative_2024,hass_extreme_2024,hass_universal_2025,hass_super-universal_2025}, which describes growth processes of random media~\cite{kardar_dynamic_1986}, turbulent liquid crystals~\cite{takeuchi_universal_2010, prolhac_height_2011}, and directed polymer~\cite{kardar_scaling_1987, halpin-healy__2012}. However, diffusion across surfaces and through volumes is physically relevant, and thus it is necessary to understand diffusion in correlated environments in $d>1$.

\textit{RWRE Model Description---}
We consider random environments in which the probability of moving in each cardinal direction changes at each time and each location, as in a random forcing field. Let $ \bm{n} = \{\hat{x},\hat{y},-\hat{x},-\hat{y}\}$ denote the set of cardinal directions. Let $\nu$ be a probability distribution on the space of probability distributions on $\bm{n}$. We denote $\mathbb{E}_\nu[\bullet]$, $\mathrm{Var}_\nu[\bullet]$ to be the expectation value and variance over all samples of $\nu$. We define the environment $\bm{\xi} = \left\{\xi_{\vec{x}, t}\, : \, \vec{x}\in\mathbb{Z}^2,t\in\mathbb{Z}_{\geq0}\right\}$ where $\xi_{\vec{x}, t}$ is a probability distribution on $\bm{n}$, independently sampled according to $\nu$, for each lattice site $(\vec{x}, t)$. We refer to $\xi_{\vec{x}, t}$ as the jump distribution of a random walk. Since every jump distribution is only defined on the cardinal directions, $\bm{n}$, we only consider nearest neighbor random walks. We also only consider environments with no net drift such that $\sum_{\hat{n}\in\bm{n}} \mathbb{E}_{\nu}\left[\xi\left(\hat{n}\right)\right]\hat{n} = \vec{0}$, or equivalently, $\mathbb{E}_\nu \left[\xi\left(\hat{n}\right)\right]=1/4$ for all $\hat{n}$ in $\bm{n}$. Note, here and below we drop the subscript on $\mathbb{E}_\nu[\xi(\bullet)]$ since all $\xi_{\vec{x}, t}$ are independent and identically distributed (i.i.d.) according to $\nu$.  

Let $\vec{S}(t) \in \Z^2$ denote a random walk starting at the origin such that $\vec{S}(0) = \vec{0}$. Let $\mathbb{P}^{\bm{\xi}}$ denote the probability measure of a random walk in the environment $\bm{\xi}$, and let $\mathbb{E}^{\bm{\xi}}[\bullet]$ and $\mathrm{Var}^{\bm{\xi}}[\bullet]$ be the corresponding expectation value and variance. In the environment $\bm{\xi}$, the probability of a random walk at $\vec{S}(t) = \vec{x}$ transitioning to $\vec{S}(t+1) = \vec{x}+\hat{n}$ is
\begin{equation}
    \label{eqn:prob}
    \mathbb{P}^{\bm{\xi}}\left(\vec{S}(t+1) = \vec{x} + \hat{n} \mid \vec{S}(t) = \vec{x}\right) = \xi_{\vec{x},t}\left(\hat{n}\right)
\end{equation}
for $\hat{n} \in \bm{n}$. Then the probability of a random walk being at site $\vec{x}$ at time $t$, $\mathbb{P}^{\bm{\xi}}\left(\vec{S}(t) = \vec{x}\right)$, obeys the recursion relation
\begin{equation}
    \label{eqn:transitionProb}
    \mathbb{P}^{\bm{\xi}}\left(\vec{S}(t+1)=\vec{x}\right) = \sum_{\hat{n}\in{\bm{n}}} \mathbb{P}^{\bm{\xi}}\left(\vec{S}(t)=\vec{x}-\hat{n}\right)\xi_{\vec{x}-\hat{n}, t}\left(\hat{n}\right)
\end{equation}
with the initial condition $\mathbb{P}^{\bm{\xi}}\left(\vec{S}(0) = \vec{0}\right) = 1$.

We define  
\begin{equation}\label{eqn:lambda}
    \lambda_{\mathrm{ext}} \coloneqq \frac{1}{2}\frac{D_{\mathrm{ext}}}{D - D_{\mathrm{ext}}}
\end{equation}
where 
\begin{equation}\label{eqn:D}
    D \coloneqq \frac{1}{2} \sum_{\hat{n}\in\bm{n}} \left(\hat{n}\cdot\hat{n}\right)\mathbb{E}_\nu \left[\xi\left(\hat{n}\right)\right]
\end{equation}
is the diffusion coefficient of a random walk in the average environment. Since we only consider environments where $\mathbb{E}_\nu[\xi(\hat{n})] = 1/4$ for all $\hat{n} \in \bm{n}$, then $D=1/2$. We define $D_\mathrm{ext}$, the \textit{extreme diffusion coefficient}, as 
\begin{align}\label{eqn:DExt}
    D_{\mathrm{ext}} \coloneqq& \frac{1}{2} \mathrm{Var}_\nu \left[\mathbb{E}^{\bm{\xi}}\left[\vec{Y}\right]\right] 
    \\
    =& \frac{1}{2}\mathbb{E}_\nu\left[\left(\mathbb{E}^{\bm{\xi}}\left[\vec{Y}\right]\right)^2\right] - \frac{1}{2}\mathbb{E}_\nu\left[\mathbb{E}^{\bm{\xi}}\left[\vec{Y}\right]\right]^2  \nonumber
\end{align}
where $\vec{Y}$ is a single step in a random walk, drawn from $\xi_{\vec{x},t}$. Thus we see $\mathbb{E}^{\bm{\xi}}\left[\vec{Y}\right]=\sum_{\hat{n}\in\bm{n}}\hat{n}\xi_{\vec{x},t}\left(\hat{n}\right)$. Since we only consider net drift free systems, $\mathbb{E}_\nu\left[\mathbb{E}^{\bm{\xi}}\left[\vec{Y}\right]\right] = \vec{0}$, so we drop the second term in Eqn.~\eqref{eqn:DExt}. Thus, $D_\mathrm{ext}$ simplifies to 
\begin{align}\label{eqn:Dext2}
    D_{\mathrm{ext}} & = \frac{1}{2} \mathbb{E}_\nu\left[\left(\sum_{\hat{n}_1\in{\bm{n}}}\hat{n}_1\xi_{\vec{x},t}\left(\hat{n}_1\right)\right)\cdot\left(\sum_{\hat{n}_2\in\bm{n}}\hat{n}_{2} \xi_{\vec{x},t}\left(\hat{n}_{2}\right)\right)\right].
\end{align}

Note that we have defined $\lambda_\mathrm{ext}$ and $D_\mathrm{ext}$ as the $d=2$ extension of their definition in studies of the $d=1$ model~\cite{hass_extreme_2024}. In the case of $d=1$, a single step of a random walk is a scalar quantity, therefore $D_\mathrm{ext} = \frac{1}{2}\mathbb{E}_\nu \left[\left(\sum_i \xi_{x,t}(i) i\right)^2\right]$. To appropriately generalize this definition to $d=2$, a single step becomes a vector, and thus we take the dot product to compute the variance. 

We now describe the choices of $\nu$ that we implement numerically. These distributions include environments in which random walks stick to one another, and environments which are homogeneous, as in the case of classical diffusion. Note that the homogeneous environment which returns classical diffusion is the one where $\xi(\hat{n}) = 1/4$ for all $\hat{n}\in\bm{n}$, called the \emph{simple symmetric random walk (SSRW)}. These choices probe the universality of $d=2$ RWRE models.
\begin{enumerate} 
    \item \emph{The Dirichlet distribution}: Let  $\xi\left(\hat{n}\right) = X_{\hat{n}} / \left(\sum_{\hat{n}} X_{\hat{n}}\right)$ for $\hat{n} \in \bm{n}$, where $X_{\hat{n}}$ are independent Gamma random variables with shape $\alpha\in\mathbb{R}_{>0}$ and rate $\beta=1$ such that $X_{\hat{n}}=x$ with probability $x^{\alpha-1}e^{-x}/\Gamma\left(\alpha\right)$ where $\Gamma(x)$ is the Gamma function. This distribution is completely determined by the parameter $\alpha$. As $\alpha\rightarrow\infty$, $\xi\left(\hat{n}\right)\rightarrow 1/4$ for all $\hat{n} \in \bm{n}$, so the SSRW is obtained.  As $\alpha\rightarrow 0$, $\xi\left(\hat{n}\right) = 1$ for one $\hat{n} \in \bm{n}$ and $\xi\left(\hat{n}\right)=0$ otherwise. This corresponds to a perfectly sticky environment so that the jump distribution is concentrated in one direction.  In this work we choose values of $\alpha$ in steps of size $\sqrt{10}$ ranging from $1/\sqrt{1000}$ to $\sqrt{1000}$.
    
    \item \emph{Log-normal}: Let $\xi\left(\hat{n}\right) = X_{\hat{n}}/\left(\sum_{\hat{n}} X_{\hat{n}} \right)$, where $X_{\hat{n}}$ are independent log-normal random variables with location $\mu=0$ and scale $\sigma=1$ such that $X_{\hat{n}}=x$ with probability $\left(e^{-\left(\ln{x}\right)^2/2}\right)/\left( x \sqrt{2\pi}\right)$.
    
    \item \emph{Random delta}: Let $\vec{X_1}$ and $\vec{X_2}$ be two random vectors uniformly sampled from $\bm{n}$ without replacement, and set $\xi\left(\vec{X_1}\right)=\xi\left(\vec{X_2}\right)=1/2$ and $\xi(\hat{n}) = 0$ otherwise. 
    
    \item \emph{Corner}: Let $X_1$ and $X_2$ be independent uniform random variables on the interval $[0,1]$, and set $\xi\left(\hat{y}\right) = X_1/2, \,\xi\left(\hat{x}\right) = \left(1-X_1\right)/2$ and similarly,  $\xi\left(-\hat{x}\right) = X_2/2, \,\xi\left(-\hat{y}\right) = \left(1-X_2\right)/2$. 
\end{enumerate}

We study the statistics of the tail probability, $\pMeas$ where $|\vec{x}|$ is the $L^2$ norm of the vector $\vec{x}$. This is the probability that a random walk in a given random environment is past a circle with radius $r$. A visualization of the definition of the tail probability is shown in Figure~\ref{fig:model} (bottom). The tail probability is relevant to the behavior of extreme particles, and its $d=1$ analogue has been studied extensively for the RWRE model~\cite{barraquand_random-walk_2017, le_doussal_diffusion_2017, barraquand_moderate_2020, parekh_hierarchy_2024,das_multiplicative_2024, hass_universal_2025}. As in those works, we look at the statistics of $\ln{ \left(\pMeas \right)} $.

\textit{Numerical Methods and Results---}
To generate pairs of radii and times, we evaluate different functional forms of $r$. We probe a wide array of pairs of $r$ and $t$, as described in 
the end matter. 

We numerically evolve the probability mass function of a random walk by direct implementation of Eqn.~\eqref{eqn:transitionProb} for a $(t+1) \times (t+1)$ lattice with origin in the center out to time $t=10000$. We choose this lattice size to optimize the memory usage of our high performance computing cluster. As this will necessitate probability mass leaving our simulation volume, we use absorbing boundary conditions to collect this probability mass. The effects of the absorbing boundary condition on the tail probability are negligible. We average our data over 500 realizations of $\bm{\xi}$ for each $\nu$.

\begin{figure}
    \centering
    \includegraphics[width=\columnwidth]{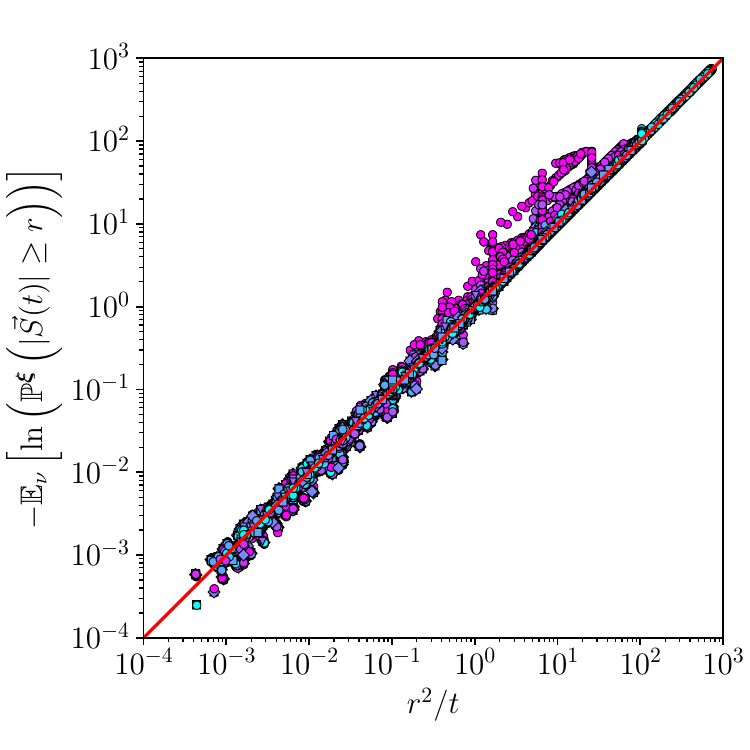}
    \caption{Collapse of the negative mean, with respect to $\nu$, of the natural log of the tail probability to a master curve when plotted against the Gaussian prediction $r^2/t$. Eqn.~\eqref{eqn:meanPrediction} is plotted as a red line, showing the validity of this prediction.}
    \label{fig:mean}
\end{figure}

As shown in Fig.~\ref{fig:mean}, the mean of the log of the tail probability, $\mathbb{E}_\nu \left[\ln{\left(\pMeas\right)}\right]$, yields the first order Gaussian behavior as $t \rightarrow \infty$ such that~\cite{le_doussal_diffusion_2017}
\begin{equation}
    \mathbb{E}_\nu \left[\ln{\left(\pMeas\right)}\right] \approx -\frac{r^2(t)}{t}. 
    \label{eqn:meanPrediction}
\end{equation}
We expect this Gaussian behavior because the average environment is homogeneous, as in classical diffusion.

\begin{figure}
    \centering
    \includegraphics[width=\columnwidth]{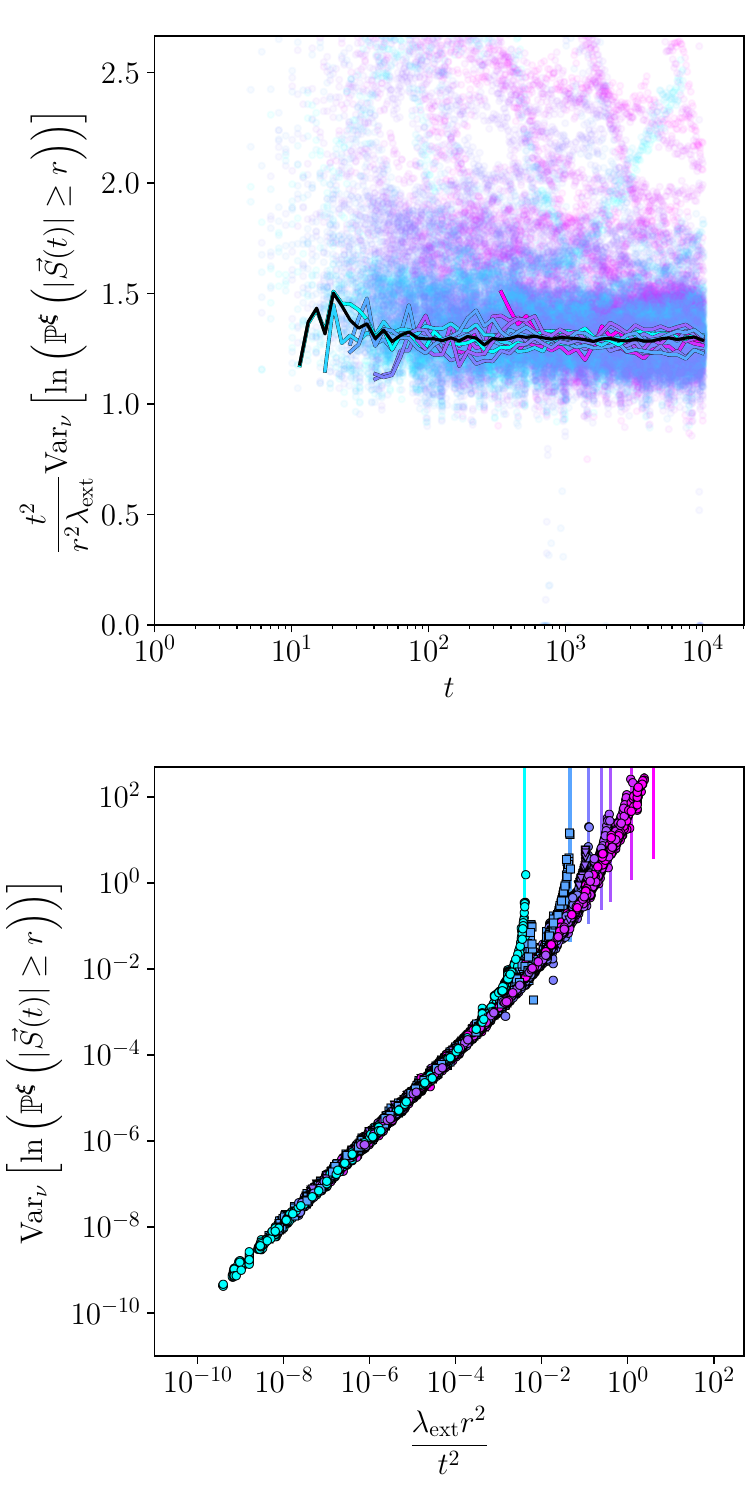}
    \caption{Universal critical behavior of the tail probability in the large deviation regime. Color gradient encodes the value of $\lambda_\mathrm{ext}$ from 0.008 (blue) to 8 (pink). Top: Variance of $\ln{\left(\pMeas\right)}$ as a function of $t$, normalized by $\lambda_\mathrm{ext}r^2(t)/t^2$.  The black line plots the median across all $\nu$, calculated as described in the text. The solid colored lines plot the median for each $\nu$. Bottom: Variance of $\ln{\left(\pMeas\right)}$ as a function of $\lambda_\mathrm{ext} r^2(t)/t^2$. Markers indicates specific choices of $\nu$: Dirichlet (circles), random-delta (triangles), and the corner distribution (squares). Vertical lines indicate where $r=t$ for each value of $\lambda_\mathrm{ext}$.}
    \label{fig:collapsedData}
\end{figure}

In Fig.~\ref{fig:collapsedData} (top) we plot the scaled variance $\frac{t^2}{r^2(t) \lambda_\mathrm{ext}}\mathrm{Var}_\nu \left[\ln{\left(\pMeas\right)}\right]$ for every $r, t,$ and $\nu$ as a function of time, $t$. We plot the median value of the scaled variance in each of 50 logarithmically spaced bins in time, for each choice of $\nu$, colored by the value of $\lambda_\mathrm{ext}$.  We plot the median value of all data in black.  The scaling factor $t^2 / r^2 \lambda_\mathrm{ext}$ collapses nearly all of the data onto a single constant as a function of time. Thus, the large deviation regime in which $r \propto t$ will result in a constant-in-time variance of the tail probability. Therefore, the critical scaling regime is the large deviation regime. As discussed below and in the end matter, when $r/t\rightarrow 1$, there is non-universal, but easily explained, behavior. This data is excluded from the median calculation. 

We find that the data collapses onto
\begin{equation}\label{eqn:collapse}
    \mathrm{Var}_\nu \left[\ln{\left(\pMeas\right)}\right] = c \lambda_\mathrm{ext} \frac{r^2}{t^2}
\end{equation}
for $r < t$, where we empirically determine $c \approx 1.3$. Note that for $d=1$ RWRE models this prefactor is analytically determined~\cite{hass_extreme_2024}. We re-plot our data in Fig. \ref{fig:collapsedData} (bottom) onto this universal form, which shows that $\lambda_{\mathrm{ext}}$ captures all relevant information about the environment (i.e. the choice of $\nu$).

The parameter $\lambda_\mathrm{ext}$ depends on the stickiness of random walks in the given environment, i.e. how likely two random walks at the same site and same time will remain together. To illustrate the effect of $\lambda_\mathrm{ext}$, consider the following limits. As $\lambda_\mathrm{ext}\rightarrow 0$, the jump distribution $\xi_{\vec{x},t}$ approaches classical diffusion where there is no correlation in the movement of particles at the same site $(\vec{x}, t)$ and thus the variance will approach zero. In contrast, as $\lambda_\mathrm{ext}\rightarrow\infty$, the jump distribution approaches completely sticky behavior, so particles coalesce at the same site $(\vec{x}, t)$ and the variance will approach infinity. 

For a given value of $\lambda_\mathrm{ext}$, as $r$ approaches $t$ we see the transition to non-universal behavior. When $r=t$, averaging only occurs over the 4 cardinal direction paths, as opposed to the many possible paths when $r$ is substantially less than $t$. In the absence of correlated averaging the variance of the tail probability scales linearly with $t$, as shown in the end matter. When plotted on our master curve, this appears as a deviation to a vertical line at $\lambda_\mathrm{ext}$, as illustrated in Fig.~\ref{fig:collapsedData} (bottom).

\textit{Conclusion---}
We have shown in this paper that RWRE models exhibit universal fluctuations of the tail probability in $d=2$. We find the critical scaling occurs for $r\propto t$, in contrast to the $d=1$ case for which the critical scaling regime is $r \propto t^{3/4}$. As the large deviation regime is the upper bound on scalings for our model, this means that $d=2$ RWRE models exhibit universal behavior in \textit{every} scaling. Additionally, the fluctuations of the tail probability decrease in time at every scaling except the critical scaling. Thus, RWRE models in $d=2$ approach classical behavior at long times in every regime except for the large deviation regime. This results raises the important question of whether or not a critical regime exists in $d>2$. 

Drillick and Parekh conjecture a critical scaling regime for $d=2$ RWRE models~\cite{drillick_random_2025} that differs from the empirically observed large deviation regime. In that work, by studying the tail probability past a line, the authors find the critical scaling regime to occur at line positions $\propto t / \sqrt{\ln{t}}$. This discrepancy is intriguing and will require more work. The fluctuations in the tail probability past a line can be understood as effectively local since they will be dominated by the fluctuations at the point of the line's closest approach to the origin. However, the tail probability past a circle will always be global since every point on the circle is equidistant to the origin and thus the fluctuations are not dominated by any single point. One might expect these two measurements to differ by logarithmic corrections. Further work is necessary to resolve this difference: empirical studies of the fluctuations past a line and analytical studies of the fluctuations past a circle.

The critical scaling regime in $d=1$ relates to the $(1+1)d$ KPZ equation. As we have uncovered the $d=2$ critical scaling regime, it is natural to ask whether it relates to the $(2+1)d$ KPZ equation~\cite{halpin-healy__2012,halpin-healy_extremal_2013}. Perhaps a detailed examination of the full distribution of the tail probability will be illuminating. Additionally, it would be valuable to extend the exploration of $d=2$ RWRE models to cover long range jump kernels as well as other measurements such as probability to reach a point and probability to reach a line.

\textit{Acknowledgements---}
This work was funded by the W.M. Keck Foundation Science and Engineering grant on ``Extreme Diffusion." We also thank Axel Saenz Rodriguez, Ivan Corwin, and Hindy Drillick for discussion. The authors acknowledge Research Advanced Computing Services (RACS) at the University of Oregon for providing computing resources that have contributed to the research results reported within this publication.

\bibliographystyle{unsrtnat}
\bibliography{bib.bib}

\clearpage
\onecolumngrid
{\centering {\huge \textbf{End Matter} \par}}
\setcounter{section}{0}
\setcounter{figure}{0}
\renewcommand{\thefigure}{A\arabic{figure}}
\bigskip

\twocolumngrid

\section{Numerical Methods}
\label{appx:num}
\begin{figure}
    \centering
    \includegraphics[width=\columnwidth]{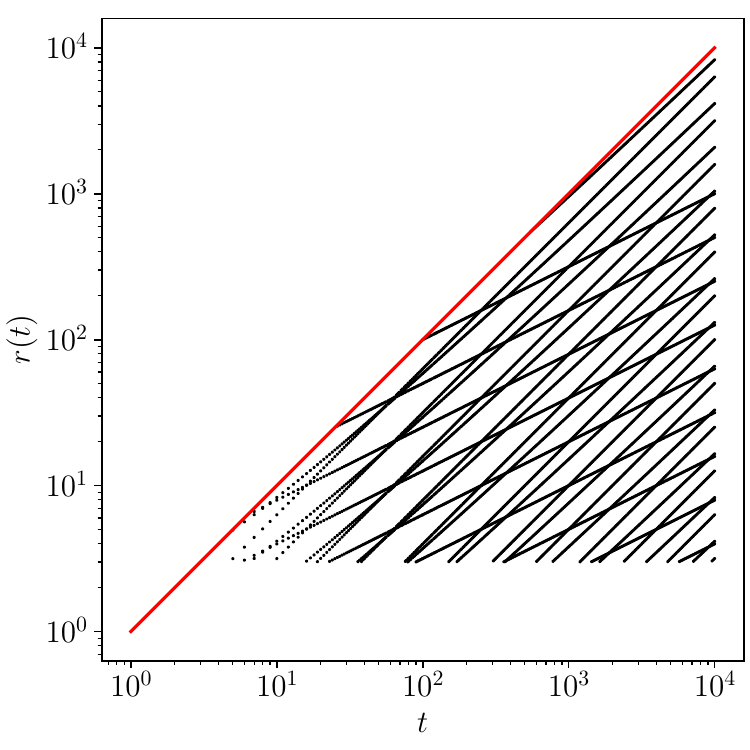}
    \caption{The set of radii, $r$, versus time, $t$, at which we measure $\pMeas$. The red line shows $r=t$.}
    \label{fig:radii}
\end{figure}

At each measurement time $t$, logarithmically spaced from $t=1$ to $t=10000$, we compute the tail probability past a number of different values for $r$. Fig.~\ref{fig:radii} plots these pairs $(r,t)$. To generate these pairs we evaluated and $r = vt^{1/2}$, $r = \frac{vt}{\sqrt{\ln{t}}}$, $r = vt$, where $v$ is varied from $10^{-3}$ to $10$.  We exclude pairs of $r$ and $t$ for which $r>t$ (shown as a red line in the figure). Additionally, we only use $r \geq 3$, to exclude short-time lattice effects. 

\section{Non-universal behavior} \label{appx:nonuniversal}
When $r/t \rightarrow 1$ we observe non universal behavior in the fluctuations of the tail probability.  This behavior is fundamentally a lattice effect driven by the narrowing of the number of possible paths between the origin and sites past $r$.  When $r=t$ there will only be 4 sites at or past $r$.  Thus, the fluctuations of $\mathbb{P}^{\bm{\xi}}(|\vec{S}(t)| = t)$ will be well described as the fluctuations of the sum of 4 random and independent paths and so our measured variance should scale linearly with $t$ for large $t$.

We explicitly compute 
\begin{equation*}
    \mathbb{P}^{\bm{\xi}}(|\vec{S}(t)| = t) = \sum_{\hat{n}\in \bm{n}} \mathbb{P}^{\bm{\xi}}\left(\vec{S}(t) = t \hat{n} \right)
\end{equation*}
as a sum over the 4 cardinal paths. 

We consider the probability of reaching site $t \hat n$ along a cardinal path,
\begin{align*}
    \mathbb{P}^{\bm{\xi}}(\vec{S}(t) = t\hat{n}) &= \prod_{i=1}^t \xi_{i\hat{n}, i}(\hat{n}), \textrm{ and} \\
    \ln\left(\mathbb{P}^{\bm{\xi}}(\vec{S}(t) = t\hat{n}) \right) &= \sum_{i=1}^t \ln\left( \xi_{i\hat{n}, i}(\hat{n}) \right).
\end{align*}
Thus, we find the variance for a single site at the end of the cardinal path to be
\begin{align*}
    \mathrm{Var}_\nu\left[ \ln\left(\mathbb{P}^{\bm{\xi}}(\vec{S}(t) = t\hat{n}) \right) \right] &= \sum_{i=1}^t \mathrm{Var}_\nu\left(\ln\left( \xi_{i
    \hat{n}, i}(\hat{n}) \right)\right) \\
    &= t \mathrm{Var}_\nu\left(\ln\left( \xi_{\vec{x}, t}(\hat{n}) \right)\right),
\end{align*}
where in the first line we used the fact that $\xi_{\vec{x}, t}$ are independent at different lattice sites $(\vec{x}, t)$ and in the second line used the property that all $\xi_{\vec{x}, t}$ are identically distributed.

To calculate the variance of the tail probability at $r = t$, we need to consider all four cardinal paths to find  
%
\begin{align*}
    &\mathrm{Var}_\nu\left[\ln\left(\mathbb{P}^{\bm{\xi}}(|\vec{S}(t)| = t)\right)\right] \\
    &=  \mathrm{Var}_\nu \left[ \ln\left(\sum_{\hat{n} \in \bm{n}} \mathbb{P}^{\bm{\xi}}\left(\vec{S}(t) = t\hat{n}\right)\right) \right] \\
    &\approx \mathrm{Var}_\nu \left[ \ln\left(4 \mathbb{P}^{\bm{\xi}}\left(\vec{S}(t) = t\hat{n}\right)\right) \right] 
\end{align*}
using the approximation
\begin{equation*}
    \sum_{\hat{n} \in \bm{n}} \mathbb{P}^{\bm{\xi}}\left(\vec{S}(t) = t\hat{n}\right) \approx 4 \mathbb{P}^{\bm{\xi}}\left(\vec{S}(t) = t\hat{n}\right)
\end{equation*}
which we expect to be reasonable in the limit that $t \rightarrow \infty$ because the variable $\mathbb{P}^{\bm{\xi}}\left(\vec{S}(t) = t \hat{n}\right)$ is independent and identically distributed for all $\hat{n} \in \bm{n}$, except at $t=0$. Therefore,
\begin{align*}
    &\mathrm{Var}_\nu\left[\ln\left(\mathbb{P}^{\bm{\xi}}(|\vec{S}(t)| = t)\right)\right] \\
    &\approx \mathrm{Var}_\nu \left[\ln(4) + \ln\left(\mathbb{P}^{\bm{\xi}}\left(\vec{S}(t) = t \hat{n} \right)\right) \right] \\
    &=  \mathrm{Var}_\nu\left[ \ln\left(\mathbb{P}^{\bm{\xi}}(\vec{S}(t) = t\hat{n}) \right) \right] \\
    &= t \mathrm{Var}_\nu \left[\ln\left(\xi_{\vec{x}, t}(\hat{n})\right)\right]
\end{align*}
Thus, the variance of the tail probability when $r = t$ scales linearly with time as $t \rightarrow \infty$. 

\end{document}